Quantifying Regional Body –wave attenuation in a seismic prone zone of northeast India

*Nilutpal Bora*, *Rajib Biswas[*]*

*Geophysical Lab, Department of Physics, Tezpur University, Tezpur-784028, Assam*

*Corresponding Author: rajib@tezu.ernet.in*

*Abstract*

We evaluated the body wave attenuation parameter in Kopili region of northeast India. Using the modified algorithm of coda normalization method, we delineated frequency-dependent attenuation for both P and S waves. Taking more than 300 seismograms as input, we comprehensively studied microearthquake spectra in the frequency range of 1.5 to 12 Hz. The estimated values of $Q_P^{-1}$ and $Q_S^{-1}$ show strong frequency dependence. Based on this, we formulated empirical relationships corresponding to $Q_P^{-1}$ and $Q_S^{-1}$ for the study region. The relationships emerge to be $Q_P^{-1} = (23.8 \pm 6) \times 10^{-3} f^{(-1.2 \pm 0.008)}$ and $Q_S^{-1} = (10.2 \pm 2) \times 10^{-3} f^{(-1.3 \pm 0.02)}$, respectively. The ratio $Q_P^{-1}/Q_S^{-1}$ is found to be larger than unity for the entire frequency band which implies profound seismic activity and macroscale heterogeneity prevailing in the region. The study may act as the building block towards determination of source parameter and hazard related studies in the region.

1. Introduction

Attenuation parameter plays a vital role so far the propagation of seismic waves through earth medium is concerned. It is represented by the inverse of a dimensionless quantity *Q*, known as Quality factor, which refers to decay in amplitudes of seismic waves during its passage from source to the receiver site. A region with high *Q* value is generally supposed to be a stable region whereas low *Q* value represents seismically active region. In general, the main factors responsible for attenuation are geometrical spreading, transformation of seismic energy into heat and redistribution of energy due to heterogeneities present in the medium. The knowledge of attenuation parameter or its inverse ($Q^{-1}$) help in better understanding of the physical state of the Earth's interior structure (Sato, 1992). Besides, it leads to quantitative prediction of strong ground motion from the perspective of engineering seismology. In addition, unique information about lithology, physical state, and degree of rock saturation (Toksöz and Johnston, 1981) could also be retrieved. Apart from that, economic and social consequences of earthquakes can be diminished through seismic risk analysis which requires the use of a hazard map. The hazard map allows for site zones in order to implement a suitable seismic code for buildings and infrastructure. As experts claim that the study of two physical processes; firstly the seismic sources and secondly propagation of the waves, is crucial for seismic hazard mapping, attenuation being one of the properties contributing to the latter (Morsy and Abed, 2013). As such, quality factor helps in modelling of earth structure with the application to deterministic seismic hazard assessment (Montaldo *et al.*, 2005). As stated by Boore, (2003), quality factor accompanied by geometric spreading leads to estimation of path effect which helps in understanding source mechanism and site response (Abercrombie and Leary, 1993; Abercrombie, 1995; Zeng and Anderson, 1996, Bonilla et al., 1997) in order to build ground-motion prediction equations (GMPEs) or ground-motion models (GMMs) for the seismic-hazard assessment. Due to such paramount significance, there is plentiful literatures (Kumar et.al. 2016; Castro *et. al.*, 2008; Singh *et. al.* 2012; Hazarika *et. al.* 2009) regarding quantification of $Q^{-1}$ throughout the world particularly in seismically active regions. Towards estimation of $Q^{-1}$, researchers adopt different techniques. Basically, the estimates are performed in relation to *P*- and *S*-waves, covering all sorts of earthquake spectra. It is worthwhile to note that these body waves are actually dependent on prevalent lithology whose amplitudes are, in general, dependent on frequency. While inferring regional body wave attenuation, the frequency dependence necessitate a detailed examination. In northeast region of India [NER], although there are handful of documentations (Hazarika *et. al.* 2009; Singh *et. al.* 2016; Kumar et.al. 2016), however, the detailed study on frequency dependence of attenuation parameter on regional scale are very few, entailing microearthquake spectra. So far the microearthquake spectra of our study region is concerned, there is evidence of varying spectral signature. As examined in Figure 1, for two events recorded by the same station, we could observe disparity in spectral content. In our quest to analyze these varying patterns, we adopted a modified methodology by Yoshimoto (1993) to attain attenuation parameter. In this work, we tried to figure out the attenuation profile, characterized by $Q_P^{-1}$ and $Q_S^{-1}$ of Kopili region, NER, India within the frequency range of *1.5 to 12 Hz*. Apart from analyzing the frequency dependence of seismic attenuation, we explored the possible mechanisms in the context of $Q_P^{-1}$ and $Q_S^{-1}$ and their ratio. The results are interrelated with available data of different parts of the world and thereby, implications are correlated.

## 2. Seismotectonics of Kopili Region

Northeast India is categorized as seismic zone V. It had witnessed 16 large (M ≥7.0) and two great, *1897* (*M$_S$* 8.7) (Oldham 1899) and *1950* (*M$_S$* 8.6) earthquakes (Poddar 1950). With embodiment of distinct geological units, it is known as one of the highest tectonically active area in the world. So far geological units are concerned, the Himalayan frontal arc lies to the north, the highly folded Indo-Burma mountain ranges or Burmese arc lies to the east. Similarly, there exists the Brahmaputra River alluvium in the Assam valley and the Shillong–Mikir plateau in between these two arcs. To the south, thick sediments of the Bengal basin surrounds this region (Bora *et. al.* 2013). This study region having latitude *25.50°* to *27.00°* E and longitude *92°* to *93°* N, NER, India comprises the active Kopili Fault Zone. It is worthy of mention that our study area also witnessed two major earthquakes of magnitude more than 7 ($M_S$). As per reports of Kayal *et. al*, (2010), there occurred a major earthquake of magnitude ~ *7.8*($M_S$) at a depth of *50 km* along the Kopili fault zone on *10*th of January *1869*, causing severe damages to area from Dhubri to east of Imphal, including Nagaon and Silchar region too. It is likely that the epicenter of the *1869* earthquake has been located possibly at *25.25$^0$N: 93.25$^0$E* (unpublished report). Another earthquake having epicenter at *27.5$^0$N: 93.5$^0$E* having origin [M~*7.3*] occurred in *1943*, October *23*. As reported by Kumar *et al.* (2016), this complete area grew during Mesozoic to Tertiary transition and was channeled by rivers Kalang and Kopili. Besides, a broad topographic depression has been observed along Koipili fault, without any surficial expression. The NW-SE trending Kopili fault separating the Shillong Plateau from the Mikir hills, extends in the north to the Main Central Thrust (MCT) in the Himalaya as shown Figure.2. This Kopili fault is, approximately *300 km* long and *50 km* wide, extending from the western part of Manipur upto the tri-junction of Bhutan, Arunachal Pradesh and Assam. Of late, Kayal *et al*.(2006), showed that high level of seismic activity exists within the Kopili fault and events pertaining to Kopili fault zone were characterized by a depth~*50km* below the Kopili fault (Kayal *et al*, 2010). Bhattacharya *et al.* (2008) estimated three-dimensional (*3D*) P-wave velocity (*Vp*) structure of the north-east India and estimated that the Kopili fault system extended from *20* to *30 km* depth and there was a high *Vp* structure below Mikir Hills at a depth of *40 km*. As per Kundu & Gahalaut (2013), the Kopili fault located to the east of the Shillong Plateau, was also undergoing active deformation with a slip of *2.9 ± 1.5 mm/year*. Barman *et. al.* (2016) reported that Kopili fault is a transpressional fault with a right lateral slip of *2.62 ± 0.79 mm/year* at a locking depth of *3 ± 2 km*. The Kopili fault was also found to have a normal and strike slip faulting with a dip in the north-eastern direction. The recent seismicity discovered along the fault has led to speculations that the Kopili fault is one of the most seismically active faults of the region and a major earthquake could be expected in the future [Kayal *et. al.* 2006, 2010].

## 3. Dataset

The present dataset consists of *300* digital seismograms of small to moderate earthquakes recorded by six stations [see Table 1] within an epicentral distance *~180 km*. This network of stations is operated by National Geophysical Research Institute (NGRI), Hyderabad. All the six stations are equipped with three component Broadband Seismographs with GPS synchronized timings. They were operated in continuous mode of recording. The data were sampled with a digitizing frequency of *100* samples /second. The corresponding epicentral plot is furnished in Figure 2. As seen, the events are scattered with good azimuthal coverage around the receiver sites under consideration. Such type of distribution helps in cancelation of influence of source radiation pattern to best possible extent (Negi *et al.,* 2014). Besides, the seismograms of all well-located events were carefully inspected to minimize any overlapping, or early cutoff. Further, emphasis is given to those events having signal to noise ratio greater than *~2*. Velocity model of Bhattacharya (2005), compatible for Kopili region has been utilized with a view to determining the hypocentral parameters. The depths of the events vary from *8.3* to *42 km*. The root mean square value for hypocentral parameter estimates is below *0.2s*.

In this work, we have utilized the initial portion of the *P*- and *S*-wave arrivals of the micro-earthquakes to derive spectra of *P*- and *S*-waves. As we have taken into account all nearby events, hence they are characterized by small *S-P* differences. As such, we stick to small sample lengths in order to avoid contamination of *P* portion of wave with *S*-wave portion. Afterwards, we perform baseline correction and trend line removal. While computing baseline, we have taken the initial *128* samples. Having subtracted the trend from the original record, the spectra were evaluated using Fast Fourier Transform (FFT) and later on subjected to instrumental correction. In order to do FFT, we have developed our own customized Matlab code, aided by desired sample length.

## 4. Methodology

We have adopted extended coda normalization method developed by Yoshimoto et. al. (1993). The elementary idea of this technique is that, if $A_C$ is the coda spectral amplitude, $A_S$ is the S-waves' source spectral amplitude and $A_P$ is the P-waves' source spectral amplitude, then the spectral amplitude of coda wave can be represented as

$$A_c(f, t_c) \infty A_s(f) \infty A_p(f) \qquad (1)$$

where $t_C$ is the lapse time, measured roughly from the origin time and is larger than twice the direct S-wave travel time. This technique is based on some assumptions that the scattering coefficient in the region is constant, the focal mechanisms are random and the ratio of P-wave to S-wave radiated energy is constant. Therefore, the source effects, instrument response and site amplification terms, which are common in both direct waves (P- and S-waves) and coda waves, are canceled by normalizing P- and S-spectra with coda spectra.

From the aforesaid coda normalization method, $Q_P^{-1}$ and $Q_S^{-1}$ can be attained from the seismogram of earthquakes observed at different hypocentral distances by using the following equations:

$$\ln\left[\frac{A_P(f,r)r}{A_C(f,t_C)}\right] = -\frac{\pi f}{Q_P(f)V_P}r + const(f) \qquad (2)$$

$$\ln\left[\frac{A_S(f,r)r}{A_C(f,t_C)}\right] = -\frac{\pi f}{Q_S(f)V_S}r + const(f) \qquad (3)$$

where $f$ is frequency and $t_c$ is a fixed lapse time from the origin time; $r$ is the hypocentral distance, $V_p$ is the P-wave velocity, and $V_s$ is the S-wave velocity. $A_p(f, r)$ and $A_s(f, r)$ are the direct

*P*- and *S*-wave maximum amplitudes, respectively and $A_c(f, t_c)$ is the coda spectral amplitude. Aki (1980) showed that, by using several *P* and *S* waves from events at different azimuths, the effects of source radiation pattern will be nullified which is one important assumption in the derivation of the above two equations. In this region, geometrical spreading plays a vital role as it affects the decay rate of the seismic wave amplitude (Yoshimoto *et al*., 1993). The geometrical spreading $Z(r)$ is expected to be proportional to the inverse of the hypocentral distance $r^{-\alpha}$ and the exponent $\alpha$ is the geometrical spreading parameter. As reported by Frankel *et al.* (1990), the quality factor for *S*-wave does not rely on frequency while varying the value of $\alpha$ from 0.7 to 1.3. For Kanto region, Yoshimoto (1993) also reported that $Q^{-1}$ did not change considerably within the frequency band with variation of $\alpha$ to 0.75, 1, and 1.25. He also explained that the value of $\alpha$ is 1 for earthquakes with hypocentral depth greater than the Moho. Again, Havskov *et al.* (1989) considered the value of $\alpha$ to be equal to 1.0 for body wave, 0.5 for surface wave and 0.75 for diffusive wave (Sato and Fehler, 1998). Therefore, we take a geometrical spreading exponent $\alpha$ to be unity in our study region. This gives an inverse power law with hypocentral distance of $r^{-1}$, which is frequently used for assessment of direct body wave attenuation in the crust (Chin and Aki 1991; Ishida, 1992; Yoshimoto *et al.*, 1993; Singh *et al*., 2012, Kumar *et al*., 2014, Negi *et al.,* 2014).

Applying the least-squares method to the values of the left-hand side of equations (2) and (3) against the hypocentral distance for many earthquakes, we can estimate $Q_P^{-1}$ and $Q_S^{-1}$ from linear regression lines in a single plot. It is evident from the above equations (2 & 3) that the values of $A_S$, $A_P$ are functions of both *f* and *r*. Once the values of $A_S$, $A_P$ and $A_C$ are known, the values of $Q_P^{-1}$ and $Q_S^{-1}$ can be determined for different frequency ranges.

## 5. Data analysis

We calculated the spectral amplitudes of the direct *P*- and *S*-waves in a *1.28- s* time window from the beginning of *P*- and *S*-waves separately. Then, the separated signal has been filtered by using Butterworth bandpass filter at five different frequency bands, *1~2, 2~5, 4~8, 6~12, 9~15* (Table 2) respectively. The maximum peak-to-peak amplitudes of direct *P*- and *S*-waves are measured for a *1.28 s* length window from the onset of P and S waves on the filtered seismograms. Half values of the peak-to-peak amplitudes represent $A_P(f,r)$ and $A_S(f,r)$ respectively. $A_P(f, r)$ is obtained in vertical-component record and $A_S(f, r)$ is read in NS-component since the maximum amplitudes of S wave are nearly equal in both the NS and EW components. Coda-spectral amplitude $A_C(f, t_C)$ is calculated for a *1.28-s* time window at the lapse time $t_C = T_0 + 2t_s$ which is taken as twice that of the S-wave travel time (Aki and Chouet, 1975; Rautian and Khalturin, 1978) and $T_0$ is the origin time of an earthquake. In the analysis, $A_C$ is considered as root mean square amplitude of the coda window which is taken from the N-S component of the seismogram for each frequency band. We take $V_P = 6\ km/s$ and $V_S = 3.5\ km/s$ for the study region from the velocity model estimated by Gupta *et al.* (1984). As seen from the figure 2, our events are randomly distributed around the receivers, so the effect of source radiation pattern is considered as negligible. We substituted $A_P(f,r)$, $A_S(f,r)$ and $A_C(f, t_C)$ into equations (2) and (3) and then got $Q_P^{-1}$ and $Q_S^{-1}$ by the least-squares method. The slopes of the best-fitted lines are used to estimate $Q_P^{-1}$ and $Q_S^{-1}$ using the following relations [from Eqs. (2) and (3)]

$$Slope = -\frac{\pi f}{Q_P(f)V_P}\ for\ P\ wave$$

$$Slope = -\frac{\pi f}{Q_S(f)V_S}\ for\ S\ wave$$

Taking TZR and RUP as an example, we furnished coda-normalized amplitude decay of *P*- wave and *S*-wave with hypocentral distance for different frequencies (Figures 3).

## 6. Results and discussion

We have observed that the values of $Q_P^{-1}$ and $Q_S^{-1}$ for the Kopili region decrease with increase in frequency. The values of $Q_P^{-1}$ and $Q_S^{-1}$ for all the five central frequency is for the six stations are given in the Table 3 and 4. It is noted that the values of $Q_P^{-1}$ and $Q_S^{-1}$ do not vary considerably from one station to other. For instance, the value of $Q_P^{-1}$ for the stations RUP and BKD are *(12.8±1.2)* $\times 10^{-3}$ and *(12.2±2.9)* $\times 10^{-3}$ respectively at *1.5 Hz*. Looking at this trend, we are encouraged to take the arithmetic mean of $Q_P^{-1}$ and $Q_S^{-1}$ for all the six stations. Table 5 shows the average values of the $Q_P^{-1}$ and $Q_S^{-1}$ with their standard deviations. As seen in Table 5, it is observed that the values of $Q_P^{-1}$ and $Q_S^{-1}$ decrease with rise in frequency. As per example, $Q_P^{-1}$ declines from about *(13.8±3.3)* $\times 10^{-3}$ at *1.5 Hz* to *(0.9±0.2)* $\times 10^{-3}$ at *12 Hz*. Similarly, $Q_S^{-1}$ shows diminishing trend from *(6.0±1.4)* $\times 10^{-3}$ to *(0.4±0.09)* $\times 10^{-3}$ in the same frequency range as that of $Q_P^{-1}$. All these observations suggest that there is strong dependence of quality factor on frequency.

This observed frequency dependence is analogous to those obtained in other tectonic areas such as Bhuj (India) (Padhy, 2009), Cairo Metropolitan area (Egypt) (Abdel-Fattah, 2009), Koyna (India) (Sharma *et. al.*, 2007), Kanto (Japan) (Yoshimoto *et. al.*, 1993) and NE India (Padhy and Subhadra, 2010).

Besides, the average ratio $Q_P^{-1}/Q_S^{-1}$ estimated in this study varies from *2.25* to *2.70* (see table 5) within the frequency range *1.5* to *12 Hz*. The estimated mean value is *2.6*. The ratio $Q_P^{-1}/Q_S^{-1}$ is greater than unity for the frequency range (*1.5–12 Hz*) considered in the present study.

It has been revealed in laboratory measurements that $Q_P^{-1}/Q_S^{-1}$ ratio is greater than unity in dry rocks (Toksö z *et al.*, 1979; Mochizuki, 1982; Winkler and Nur, 1982). The study of Johnston *et al.* (1979) also indicates that at surface pressure most dry rocks have $Q_P^{-1}/Q_S^{-1} > 1$. Yoshimoto *et al.* (1993) and Rautian *et al.* (1978) reported $Q_P^{-1}/Q_S^{-1}$ ratio larger than unity at frequencies higher than *1 Hz*, Similar results were also reported by Chung and Sato (2001) for South Eastern Korea, the ratio where they computed to be greater than ǂ unity for the frequency range *1.5–10Hz*. The value of ratio $Q_P^{-1}/Q_S^{-1}$ found in the present analysis is in agreement with the results of the laboratory measurements and other studies mentioned above. It is evident that $Q_P^{-1}$ is greater than $Q_S^{-1}$ by a factor of *2.42* in the frequency range *1.5-12 Hz*. This estimated value is approximately equal to the theoretical predicted ratio of *2.41* at high frequency limit for entire NE India (Sato 1984).

In order to build the frequency-dependent relationship $Q = Q_0 f^n$, we fitted the expected average values of *Q⁻¹*, as a function of central frequency by using a power law. In the expression, $Q_o$ is the quality factor at a reference frequency *f=1Hz* and *n* is the frequency-dependence coefficient, which is approximately close to unity and varies from region to region on the basis of heterogeneity of the medium (Aki 1981). As illustrated in Figure 4, average values of $Q_P^{-1}$ and $Q_S^{-1}$ are plotted against the central frequency. The standard deviations are represented by the error bars. The estimated frequency dependent formula for our study region emerges to be $Q_P^{-1} = (23.8 \pm 6) \times 10^{-3} f^{(-1.2 \pm 0.008)}$ and $Q_S^{-1} = (10.2 \pm 2) \times 10^{-3} f^{(-1.3 \pm 0.02)}$. These estimates of $Q_P^{-1}$ and $Q_S^{-1}$ correspond to those of the seismically active areas in the world.

The frequency dependencies for *P*- and *S*-waves are almost similar. We find that the attenuation for *P*-waves is greater than for *S*-waves in the entire frequency range. The obtained $Q_P^{-1}/Q_S^{-1}>1$ is observed in the upper crust of many other regions with a high degree of lateral heterogeneity (Bianco et al., 1999; Sato and Fehler, 1998).

This variation in $Q^{-1}$ values may be attributed to the heterogeneities present in the region. From the geological point of view, crustal structure of earth is endowed with heterogeneities on many scales. Aki (1980), VanEck (1988) showed a correlation between the degree of frequency dependence and the level of tectonic activity. They established higher n values for tectonically active regions as compared to that of tectonically stable regions. In our study the computed values of '*n*' are~1.2 and ~1.3 for *P*- and *S*-waves.

In order to get further insight in the observed attenuation pattern, we have reviewed the available geophysical evidences and tried to correlate them. As reported by Bhattacharya *et al.* (2002, 2008), Kopili fault is the next pronounced active zone after Shillong Plateau. They observed an intense activity along the NW-SE Kopili fault. Bhattacharya *et al.* (2002) conducted $1^0$ and $2^0$ mapping of entire NE region. They found high b-value as well as high fractal dimensions (*1.65 < D < 1.85*) along the Kopili fault. They showed that higher D values along the Kopili Fault are due to the heterogeneous transverse structure. The high b-value implies higher heterogeneities (Mogi (1962)) or presence of heterogeneous fractured rock mass. All these bear strong resemblances with our estimated values of $Q_P^{-1}$ and $Q_S^{-1}$.

Further, we have compared our results with the available worldwide data, as shown in Figure 5 and 6. As depicted in Table 6 we classified them based on the attenuation characteristics, related to the tectonic settings. The pattern of $Q_P^{-1}$ with frequency is comparable (Figure. 5) to the estimates obtained in other tectonic areas. The pattern is almost similar to that of Kummaun, India (Singh *et al.* 2012). However, the higher estimates of $Q_P^{-1}$ is mainly attributed to higher degree of heterogeneities in the crust beneath the study area. Similarly, the pattern of $Q_S^{-1}$ with frequency is comparable (Figure. 6) to the estimates obtained in Kummaun, India (Singh *et al.* 2012); Koyna, India (Sharma *et. al.* 2007). This comparison converges to the fact that the attenuation characteristics in our study area is quite comparable with the other seismically alert parts of the world.

## 7. Conclusion

In summary, we tried to estimate seismic wave attenuation in Kopili region, NER India by using direct *P*- and *S*-waves for six seismic stations in the frequency range of *1.5–12 Hz*.

The findings are:

1. Both $Q_P^{-1}$ and $Q_S^{-1}$ are frequency dependent as the values of $Q_P^{-1}$ and $Q_S^{-1}$ decrease with rising frequency from about *(13.8±3.3)*× $10^{-3}$ and *(6.0±1.4)* × $10^{-3}$ at *1.5 Hz* to *(0.9±0.2)* × $10^{-3}$ and *(0.4±0.09)* × $10^{-3}$ at *12 Hz*, respectively.

2. The frequency dependent formula for both $Q_P^{-1}$ and $Q_S^{-1}$ are obtained by using a power law fitting against the analyzed frequency range. Our estimated values of *P*- and *S*-waves attenuations are $Q_P^{-1} = (23.8 \pm 6) \times 10^{-3} f^{(-1.2\pm0.008)}$ and $Q_S^{-1} = (10.2 \pm 2) \times 10^{-3} f^{(-1.3\pm0.02)}$ respectively.

3. The estimated value of $Q_P^{-1}/Q_S^{-1}$ is *2.42* which is approximately equal to the theoretical predicted ratio (by Sato 1984) of *2.41* at high frequency limit for entire NE India.

4. The observed findings are similar to other seismically active areas like Kanto (Japan), Zarand (Iran), Koyna (India), East central Iran and Kummaun (India). The observed high attenuation is mainly attributed to high degree of heterogeneity in the crust beneath the study area.

The obtained attenuation relations may be implemented towards computation of source parameters. Additionally, the near-source simulation of earthquake ground motion of the earthquakes can also be executed by resorting to these relations for this region which will eventually help in assessing seismic hazard in the region.

**REFERENCES**

Abdel-Fattah AK (2009) Attenuation of body waves in the crust beneath the vicinity of Cairo Metropolitan area (Egypt) using coda normalization method. Geophysical Journal International. 176:126–134.

Abercrombie, R. E. (1995). Earthquake source scaling relationships from −1 to 5 M L using seismograms recorded at 2.5 km depth, J. Geophys. Res. 100, 24,015–24,036.

Abercrombie, R. E., and P. Leary (1993). Source parameters of small earthquakes recorded at 2.5 km depth, Cajon Pass, southern California: Implications for earthquake scaling, Geophys. Res. Lett. 20, 1511–1514.

Aki K, Chouet B (1975) Origin of coda waves: source, attenuation and scattering effects. Journal of Geophysical Research. 80:3322–3342.

Aki K (1980) Attenuation of shear-waves in the lithosphere for frequencies from 0.05 to 25 Hz. Physics of the Earth and Planetary Interiors 21:50–60.

Aki, K. (1981) Attenuation and scattering of short-period seismic waves in the lithosphere, Identification of seismic sources-Earthquake or Underground Explosion, eds.: Husebye,E.S., and Mykkeltveit, S., D. Reidel Publishing Co., pp. 515-541.

Barman, P., Ray, J.D., Kumar, A., Chowdhury, J.D. & Mahanta, K., (2016), Estimation of present-day inter-seismic deformation in Kopili fault zone of north-east India using GPS measurements, Geomatics, Natural Hazards and Risk, 7:2, 586-599, DOI: 10.1080/19475705.2014.983187

Bhattacharya, P.M., Majumder, R.K. & Kayal, J.R., 2002. Fractal dimension and b-value mapping in northeast India, Current Science. 82, 1486–1491.

Bhattacharya PM, Pujol J, Mazumdar RK, Kayal JR (2005) Relocation of earthquakes in the Northeast India region using joint Hypocenter determination method. Curr Sci 89(8):1404–1413

Bhattacharya, P.M., Mukhopadhyay, S., Majumdar, R.K. & Kayal, J.R., 2008. 3-D seismic structure of the northeast India region and its implications for local and regional tectonics, Journal of Asian Earth Sciences. 33, 25–41.

Bianco F, Castellano M, Del Pezzo E, Ibañez JM (1999) Attenuation of short-period seismic waves at Mt. Vesuvius, Italy. Geophysical Journal International.138:67–76

Biswas,R. Baruah, S. Bora,D. Kalita, A And Baruah, S. (2013 a), The Effects of Attenuation and Site on the Spectra of Microearthquakes in the Shillong Region of Northeast India, Pure and Applied Geophysics, Springer Basel. 170: 1833.

Biswas,R. Baruah, S. Bora,D. (2013 b), Influence of Attenuation and Site on Microearthquakes' Spectra in Shillong Region, of Northeast, India: A Case Study, Acta Geophysica, vol. 61, no. 4, Aug. 2013, pp. 886-904, DOI: 10.2478/s11600-013-0129-x

Bonilla, L. F., J. H. Steidl, G. T. Lindley, A. G. Tumarkin, and R. J. Archuleta (1997). Site amplification in the San Fernando Valley, CA: Variability of site effect estimation using the S-wave, coda, and H/V methods, Bull. Seismol. Soc. Am. 87, 710–730.

Bora, D.K., Baruah, S., Biswas, R., and Gogoi, N. K.,(2013), Estimation of Source Parameters of Local Earthquakes Originated in Shillong–Mikir Plateau and its Adjoining Region of Northeastern India, Bulletin of the Seismological Society of America. Vol. 103, No. 1, 437-446.

Boore, D. M. (2003). Prediction of ground motion using the stochastic method, Pure Appl. Geophys. 160, 635–676.

Castro, R.R., Condori, C., Romero, O., Jacques, C. & Suter, M., 2008. Seismic attenuation in northeastern Sonora, Mexico, Bulletin of the Seismological Society of America. 98, 722–732.

Chin, B. H., and Aki, K. (1991), Simultaneous study of the source, path, and site effects on strong ground motion during the 1989 Loma Prieta earthquake: a preliminary result on pervasive nonlinear site effects, Bull. Seismol. Soc. Am. 81(5), 1859–1884.

Chung TW, Sato H (2001) Attenuation of high-frequency P and S waves in the crust of southeastern South Korea. Bulletin of the Seismological Society of America. 91:1867–1874.

Frankel, A., Mcgarr, A., Bicknell, J., Mori, J., Seeber, L., and Cranswick, E. (1990), Attenuation of high frequency shear waves in the crust: measurements from New York State, South Africa, and Southern California. J. Geophys. Res.-Sol. Ea. 95(B11), 17441–17457.

Gupta, H.K., Singh, S.C., Dutta, T.K. & Saikia, M.M., 1984. Recent investigations of North East India seismicity, in *Proceedings of the International Symposium on Continental Seismicity and Earthquake Prediction,* pp. 63–71, eds Gongu, G., Xing-Yuan, M., Seismological Press, Beijing.

Havskov J, Malone S, Mcclurg D, Crosson R (1989) Coda Q for the state of Washington. Bull Seismol Soc Am 79:1024–1038.

Hazarika, D., Baruah, S., Gogoi, N.K, (2009), Attenuation of coda waves in the Northeastern Region of India; J Seismol, 13:141–160 DOI 10.1007/s10950-008-9132-0.

Ishida, M. (1992), Geometry and relative motion of the Philippine Sea plate and Pacific plate beneath the Kanto-Tokai district, Japan, J. Geophys. Res.-Sol. Ea. 97, 489–513.

Johnston, D.H., Toksoz, M.N., and Timur, A. (1979), Attenuation of seismic waves in dry and saturated rocks:I. Mechanics, Geophysics 44, 691–711.

Kayal, J.R., Arefiev, S.S., Barua, S., Hazarika, D., Gogoi, N., Kumar, A., Chowdhury, S.N., Kalita, S., 2006. Shillong Plateau earthquakes in Northeast India region: complex tectonic model. Current Science. 91 (1), 109–114.

Kayal, J.R., Arefiev, S.S., Baruah, S., Tatevossian, R., Gogoi, N., Sanoujam, M., Gautam, J.L., Hazarika, D., And Borah, D. (2010), The 2009 Bhutan and Assam felt earthquakes (Mw 6.3 and 5.1) at the Kopili fault in the northeast Himalaya region, Geomatics, Natural Hazards and Risk. 1(3), 273- 281

Kundu B, Gahalaut VK. (2013). Tectonic geodesy revealing geodynamic complexity of the Indo-Burmese arc region, north east India. Curr Sci. 104:920-933.

Kumar R, Gupta, S.C., Singh, S.P., and Kumar, A., (2016 a), The Attenuation of High-Frequency Seismic Waves in the Lower Siang Region of Arunachal Himalaya: $Q\alpha$, $Q\beta$, $Qc$, $Qi$, and $Qs$, Bulletin of the Seismological Society of America, Vol. 106, No. 4, pp-, doi: 10.1785/0120150113

Kumar, N., Mate, S., Mukhopadhyay, S. (2014), Estimation of Qp and Qs of Kinnaur Himalaya, J. Seismol. 18(1), 47–59.

Kumar Devender, Reddy D.V., Pandey Anand K. (2016 b), Paleoseismic investigations in the Kopili Fault Zone of North East India: Evidences from liquefaction chronology, Tectonophysics. 674, 65-75.

Ma'hood M, Hamzehloo H, Doloei GJ (2009) Attenuation of high frequency P and S waves in the crust of the East-Central Iran. Geophysical Journal International.179:1669–1678.

Mochizuki, S. (1982), Attenuation in partially saturated rocks, J. Geophys. Res. 87, 8598–8604.

Mogi, K., 1962. Magnitude-frequency relationship for elastic shocks accompanying fractures of various materials and some related problems in earthquakes. Bulletin of the Earthquake Research Institute. University of Tokyo, 40: 831-883.

Montaldo, V., Faccioli, E., Zonno, G., Akinci, A., Malagnini, L., 2005. Treatment of ground-motion predictive relationships for the reference seismic hazard map of Italy. J. Seismol. 9, 295–316 (Full Text via CrossRef|View Record in Scopus|Cited By in Scopus (4)).

Morsy, M. A. and Abed, A. M., 2013. Attenuation of seismic waves in central Egypt. NRIAG Journal of Astronomy and Geophysics (2013) 2, 8–17

Negi, S. S., Paul, A., Joshi, A. and Kamal, 2014. Body wave crustal attenuation in the Garhwal Himalaya, India Pure Appl. Geophysics, Springer Basel.

Oldham, R.D. (1899), Report on the great earthquake of 12th June 1897. Mem Geological Survey of India. 29, 1-379.

Padhy S (2009) Characteristics of body wave attenuations in the Bhuj crust. Bulletin of the Seismological Society of America. 99:3300–3313

Padhy, S. & Subhadra, N., 2010. Attenuation of high-frequency seismic waves in northeast India, Geophysical Journal International., 181, 453–467.

Poddar, M.C., 1950. The Assam earthquake of 15th August 1950. Indian Miner. 4,167–176.

Qin-cai, W., Jie, L., Si-hua, Z. & Zhang-li, C., 2005. Frequency-dependent attenuation of P and S waves in Yunnan region, Acta Seismologica Sinica. 18, 632–642.

Rautian TG, Khalturin VI (1978) The use of coda for determination of the earthquake source spectrum. Bulletin of the Seismological Society of America. 68:923–948

Rautian TG, Khalturin VI, Martynov VG, Molnar P (1978) Preliminary analysis of the spectral content of P and S waves from local earthquakes in the Garm, Tadjikistan region. Bulletin of the Seismological Society of America. 68:949–971

Sato H, Fehler M (1998). Seismic wave propagation and scattering in the heterogeneous earth. (AIP)American Institute of Physics. Press/Springer, New York, p 308

Sato H. (1992). Thermal structure of the mantle wedge beneath Northeastern Japan: Magmatism in an island arc from the combined data of seismic anelasticity and velocity and heat flow. Journal of Volcanology and Geothermal Research. 51: 237~252.

Sato, H., 1984. Attenuation and envelope formation of three-component seismograms of small local earthquakes in randomly inhomogeneous lithosphere, Journal of Geophysical Research. 89, 1221–1241.

Sharma B, Teotia SS, Kumar D (2007) Attenuation of P, S, and coda waves in Koyna region, India. Journal of Seismology. 11:327–334.

Sharma, B., Gupta, K.A., Devi, K.D., Kumar, D., Teotia S.S. & Rastogi, B.K., 2008. Attenuation of High-Frequency Seismic Waves in Kachchh Region, Gujarat, India,. *Bull. seism. Soc Am.*, 98(5), 2325–2340.

Singh C, Singh A, Bharathi VKS, Bansal AR, Chadha RK (2012) Frequency-dependent body wave attenuation characteristics in the Kumaun Himalaya. Tectonophysics. 524:37–42

Singh, S., Singh. C., Biswas. R., Mukhopadhyay. S. and Sahu. H. (2016) Attenuation characteristics in eastern Himalaya and southern Tibetan Plateau: An understanding of the physical state of the medium, Physics of earth and planetary Interiors, 257: 48–56

Toksö Z, M.N., Johnston, A.H., and Timur, A. (1979), Attenuation of seismic waves in dry and saturated rocks—I. Laboratory measurements, Geophysics 44, 681–690.

Toksöz MN, Johnston DH (1981) Preface. In: Toksöz MN, Johnston DH (eds) Seismic wave attenuation. Society of Exploration Geophysicists, Tulsa, pp v–vi

Van Eck, T. (1988), Attenuation of coda waves in Dead Sea region, Bulletin of the Seismological Society of America. 78, 770–779.

Winkler, K.W. and Nur, A. (1982), Seismic attenuation effects of pore fluids and frictional sliding, Geophysics 47, 1–15.

Yoshimoto K, Sato H, Ohtake M (1993) Frequency-dependent attenuation of P and S waves in the Kanto Area, Japan, based on the coda-normalization method. Geophysical Journal International.114:165–174.

Yoshimoto K, Sato H, Ito Y, Ito H, Ohminato T, Ohtake M (1998) Frequency-dependent attenuation of high-frequency P and S waves in the upper crust in western Nagano, Japan. Pure and Applied Geophysics. 153:489–502.

Zeng, Z., and J. G. Anderson (1996). A composite source model of the 1994Northridge earthquake using genetic algorithms, Bull. Seismol. Soc.Am. 86, 71–83.

**Lists of Figures**

**Figure** 1. Displacement spectra of the nearby event with respect to the farthest event recorded at TZR-station, NER, India

**Figure** 2. Map showing major tectonic features of Kopili Fault region [after Biswas *et. al.*, (2013) a, b.]. These are as indicated: Main Central Thrust, Main Boundary Thrust, Kopili fault, Dauki fault, Oldham fault (OF), Borapani Shear Zone (BS). The filled circles represent the epicenters used for our study. The stations temporarily deployed in Kopili region are indicated by Triangles. The star symbols represent the Great and major earthquakes originating in and around Shillong Plateau (Inset map of India).

**Figure** 3: Plots of coda-normalized P and S-wave amplitudes obtained using the N–S component of the seismograms versus the hypocentral distance for each frequency band for station (a) TZR and (b) RUP. The estimated $Q_P^{-1}$ and $Q_S^{-1}$ values are also shown.

**Figure** 4: Plot of mean values of $Q_P^{-1}$ and $Q_S^{-1}$ for Kopili region. Bars show standard error from mean at each frequency. The frequency dependent equation is also given by using power law fitting.

**Figure** 5: Comparison of $Q_P^{-1}$ values obtained from our study area with other parts of the world having different tectonic setting.

**Figure** 6: Comparison of $Q_S^{-1}$ values obtained from our study area with other parts of the world having different tectonic setting.

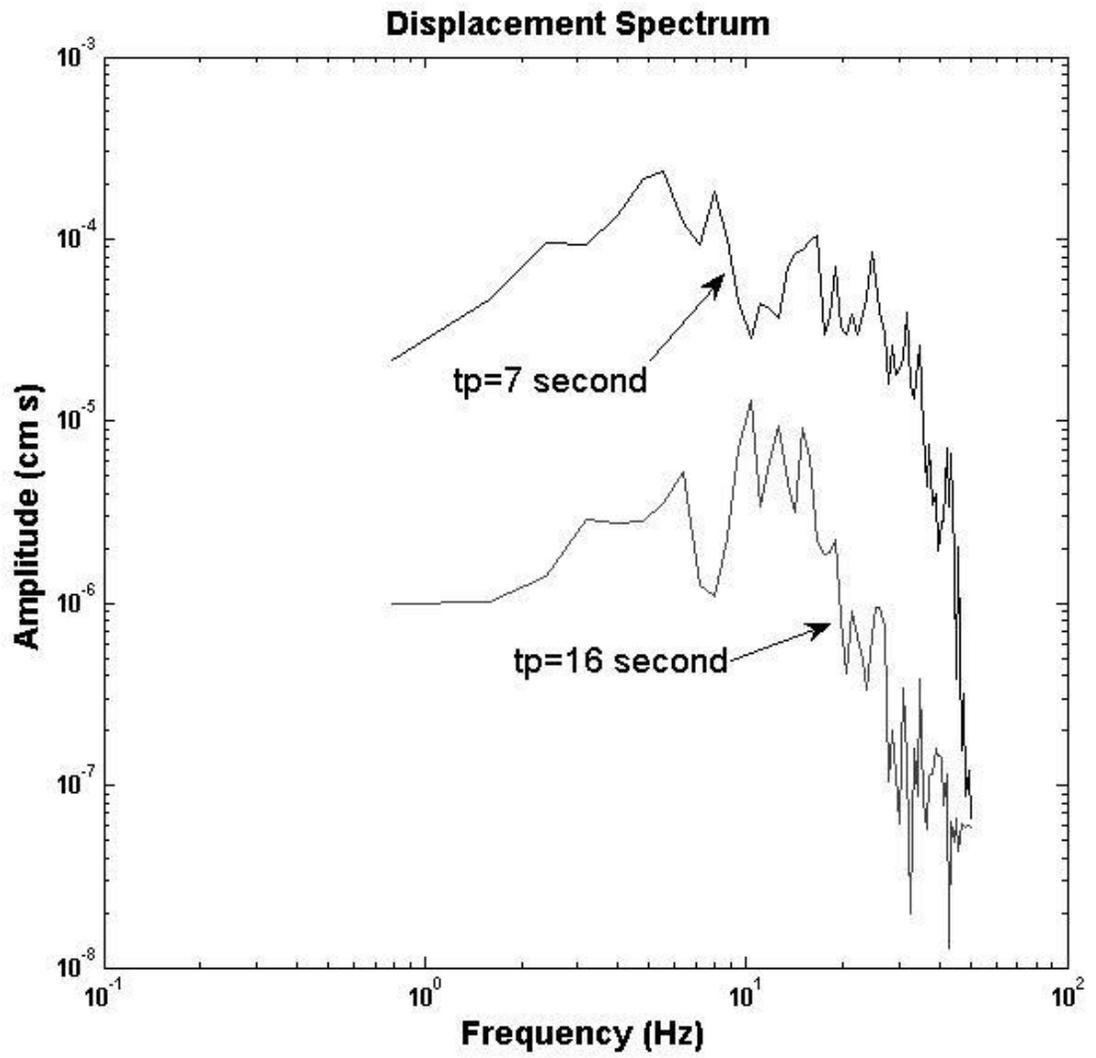

**Figure** 1.

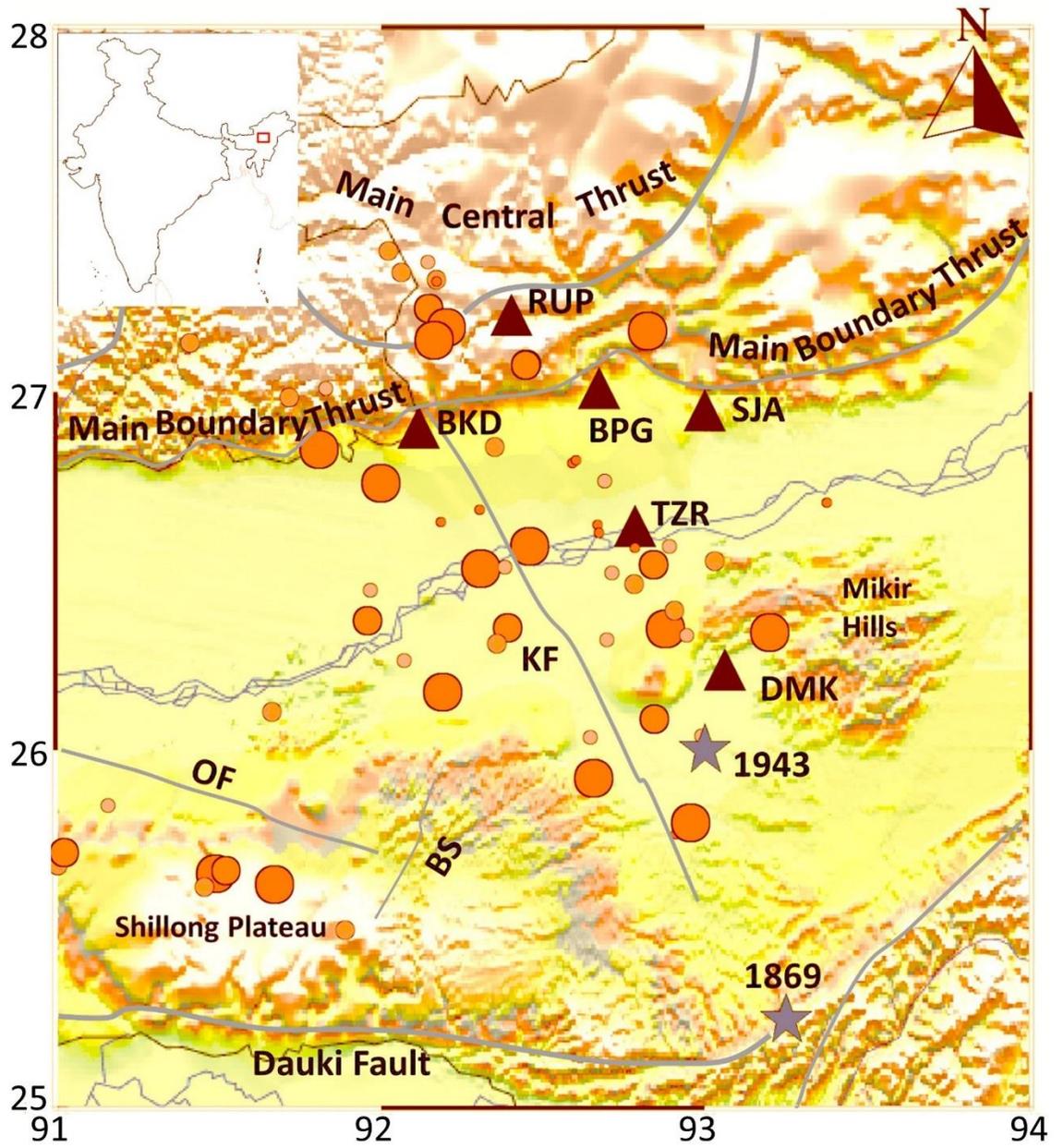

**Figure** 2.

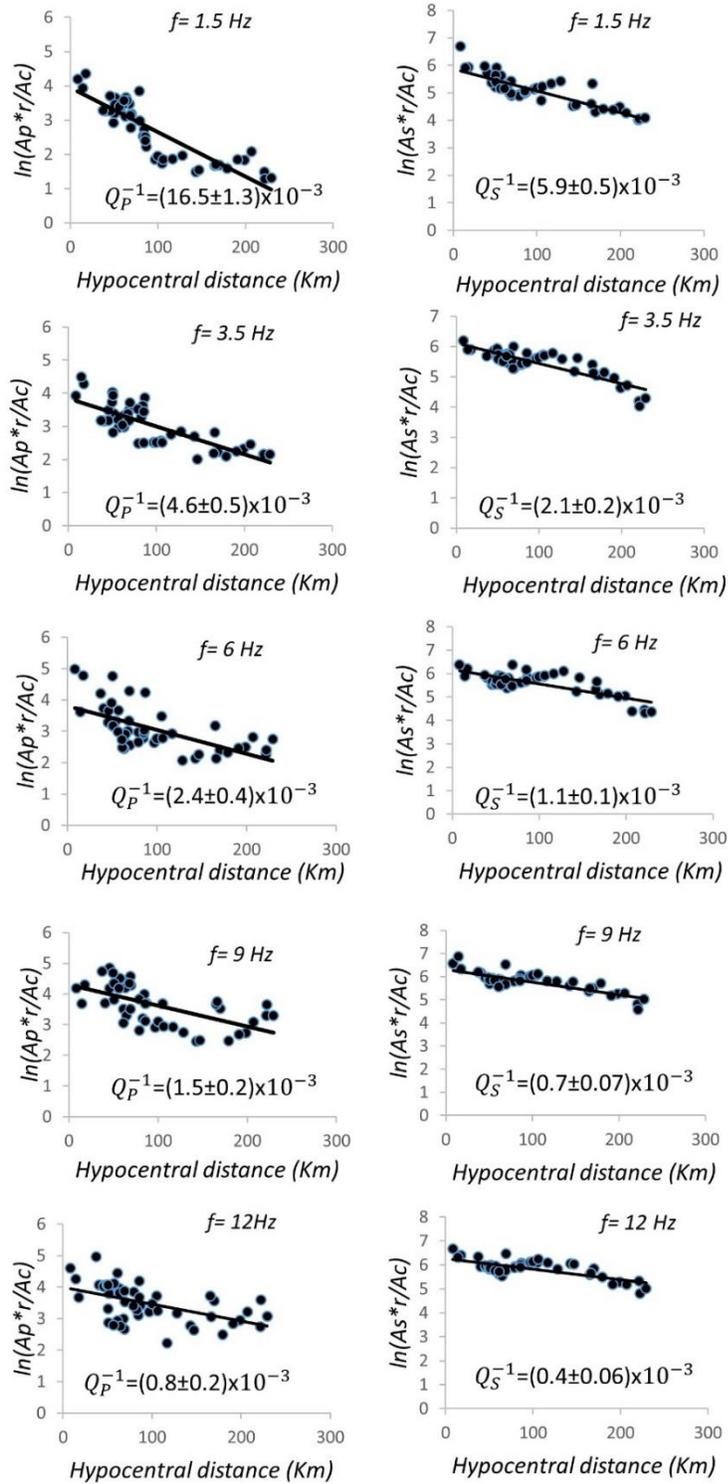

**Figure** 3:

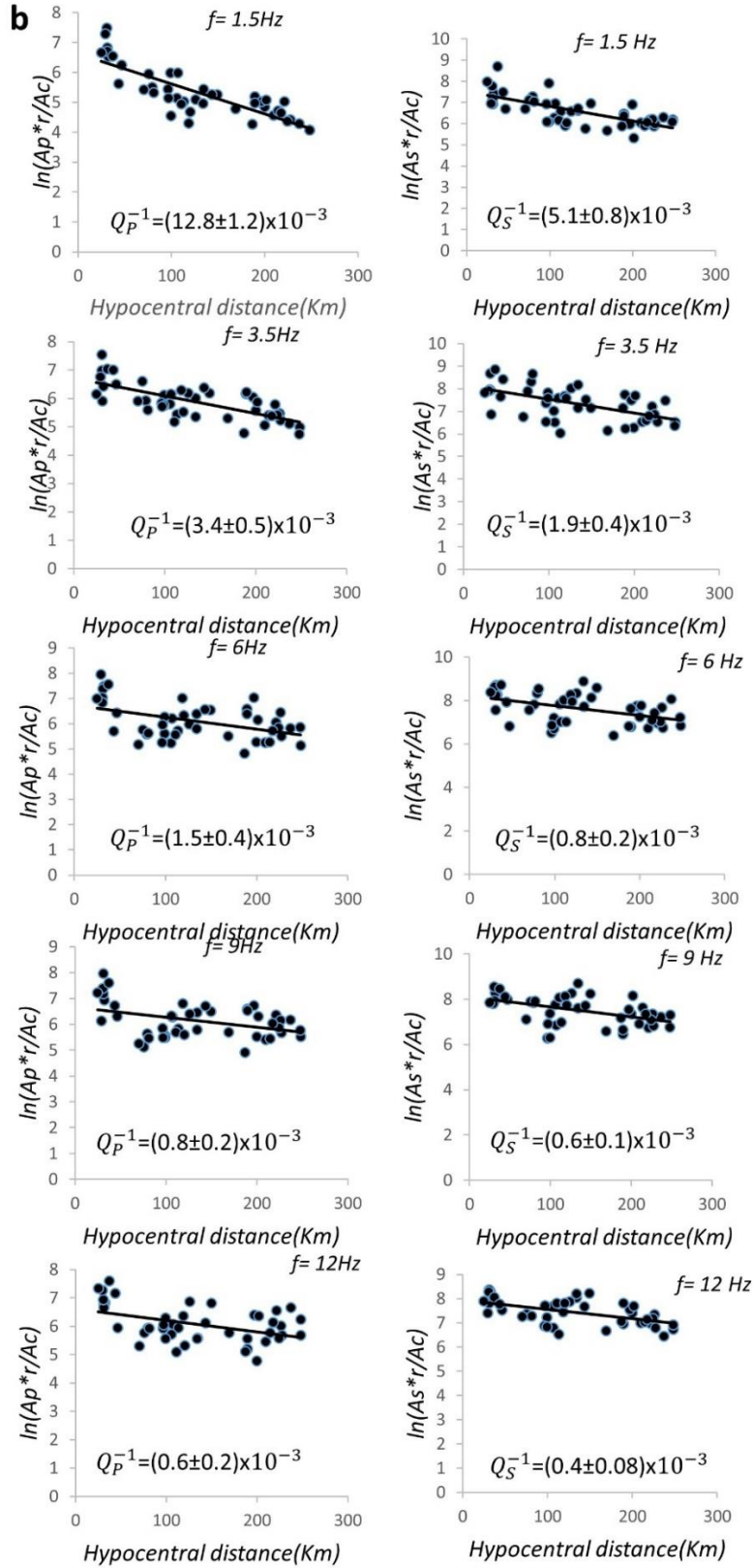

**Figure** 3: (continued)

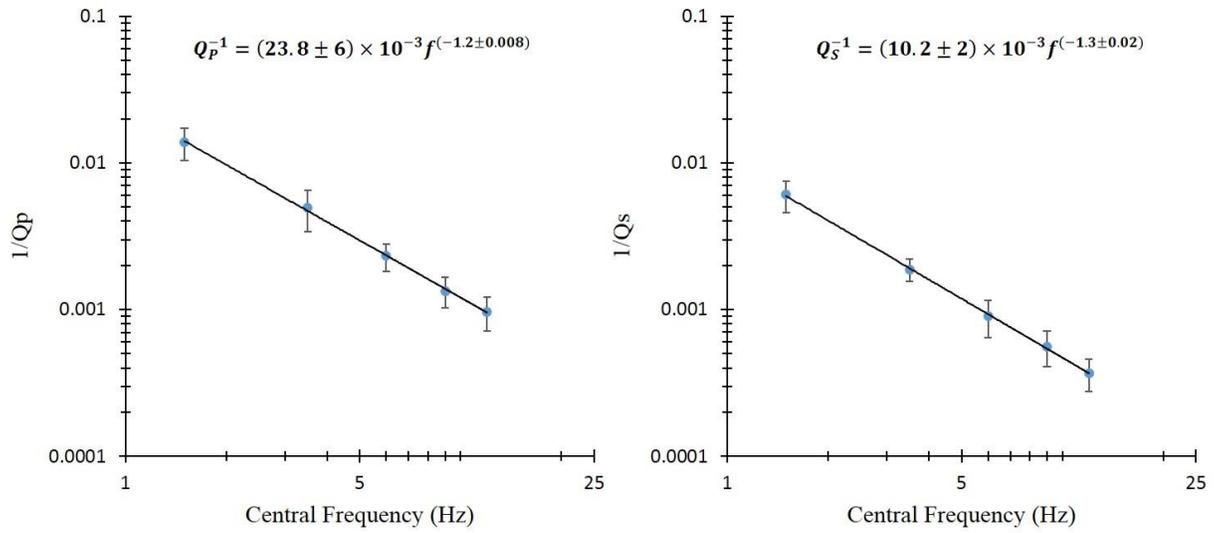

**Figure** 4:

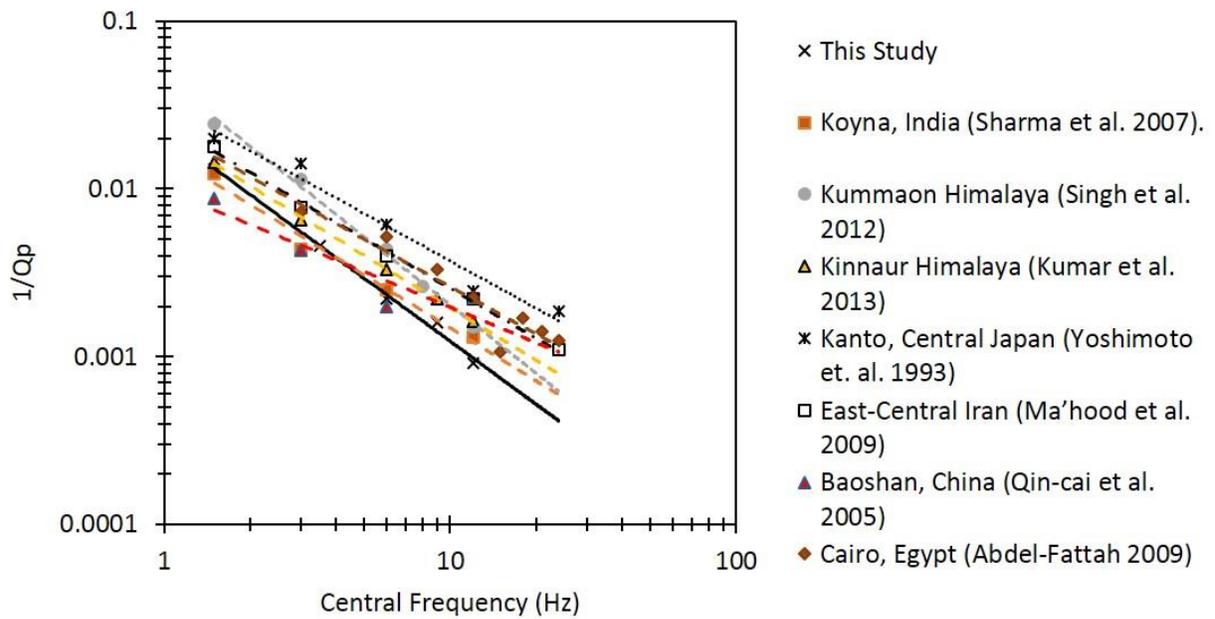

**Figure** 5:

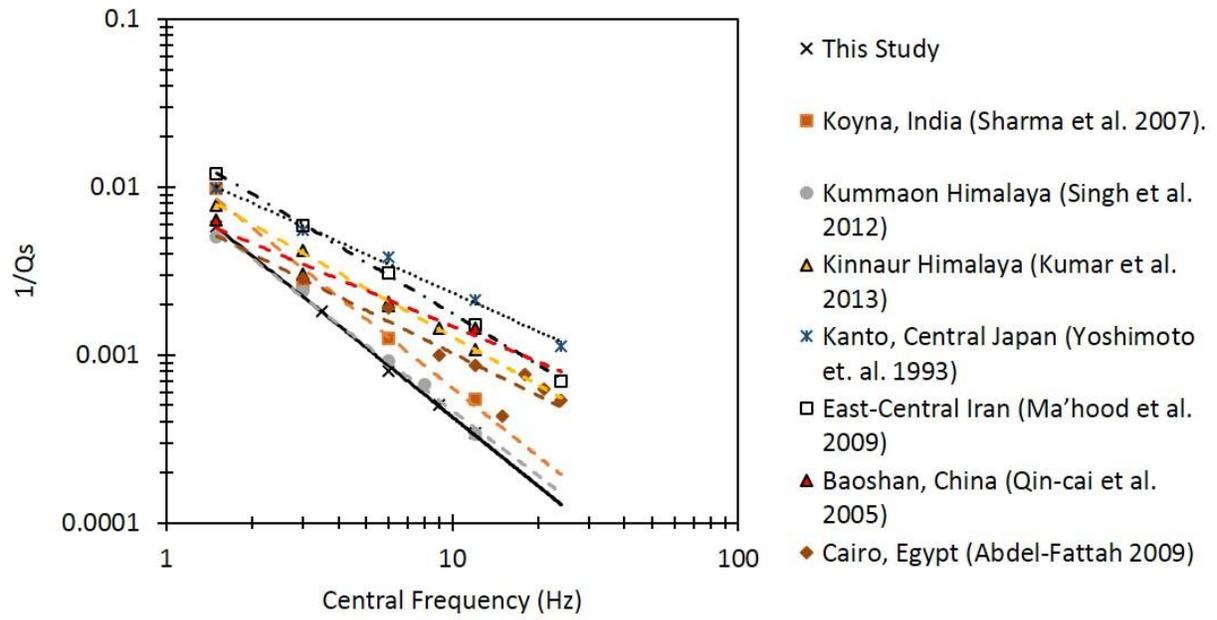

**Figure** 6:

**Lists of Tables**

**Table 1:**

| Station Name | Latitude (Degrees) | Longitude (Degrees) | Elevation (Meter) | Ground Type |
|---|---|---|---|---|
| BKD | 26.89 | 92.11 | 210 | Hard Rock |
| BPG | 26.99 | 92.67 | 130 | Hard Rock |
| DMK | 26.21 | 93.06 | 200 | Alluvium |
| RUPA | 27.20 | 92.40 | 1470 | Hard Rock |
| SJA | 26.93 | 92.99 | 150 | Hard Rock |
| TZR | 26.61 | 92.78 | 140 | Hard Rock |

**Table 2:**

| Low cutoff(Hz) | Central Frequency(Hz) | High cutoff(Hz) |
|---|---|---|
| 1 | 1.5 | 2 |
| 2 | 3.5 | 5 |
| 4 | 6 | 8 |
| 6 | 9 | 12 |
| 9 | 12 | 15 |

**Table** 3:

| Central Frequency (Hz) | TZR $Q_P^{-1}/10^{-3}$ | RUP $Q_P^{-1}/10^{-3}$ | BKD $Q_P^{-1}/10^{-3}$ | BPG $Q_P^{-1}/10^{-3}$ | DMK $Q_P^{-1}/10^{-3}$ | SJA $Q_P^{-1}/10^{-3}$ |
|---|---|---|---|---|---|---|
| 1.5 | (16.5±1.3) | (12.8±1.2) | (12.2±2.9) | (10.3±2.5) | (19.2±3.5) | (19.4±4.5) |
| 3.5 | (4.6±0.5) | (3.4±0.5) | (3.8±1.1) | (5.3±1.6) | (8.0±1.4) | (4.7±2.9) |
| 6 | (2.4±0.4) | (1.5±0.4) | (2.6±0.7) | (2.7±1.0) | (2.7±0.9) | (1.9±1.8) |
| 9 | (1.5±0.2) | (0.8±0.2) | (1.7±0.6) | (1.3±0.6) | (1.5±0.7) | (1.2±1.4) |
| 12 | (0.8±0.2) | (0.6±0.2) | (1.3±0.5) | (1.0±0.5) | (1.2±0.7) | (0.8±0.2) |

**Table** 4:

| Central Frequency (Hz) | TZR $Q_S^{-1}/10^{-3}$ | RUP $Q_S^{-1}/10^{-3}$ | BKD $Q_S^{-1}/10^{-3}$ | BPG $Q_S^{-1}/10^{-3}$ | DMK $Q_S^{-1}/10^{-3}$ | SJA $Q_S^{-1}/10^{-3}$ |
|---|---|---|---|---|---|---|
| 1.5 | (5.9±0.5) | (5.1±0.8) | (5.3±1.4) | (5.4±0.2) | (9.0±0.1) | (6.0±0.5) |
| 3.5 | (2.1±0.2) | (1.9±0.4) | (1.8±0.5) | (1.9±0.6) | (2.0±0.3) | (1.3±0.2) |
| 6 | (1.1±0.1) | (0.8±0.2) | (0.9±0.3) | (1.0±0.2) | (1.0±0.2) | (0.4±0.1) |
| 9 | (0.7±0.07) | (0.6±0.1) | (0.6±0.2) | (0.7±0.1) | (0.6±0.2) | (0.3±0.08) |

| 12 | (0.4±0.06) | (0.4±0.08) | (0.5±0.1) | (0.4±0.1) | (0.4±0.2) | (0.2±0.04) |

**Table** 5:

| Central Frequency (Hz) | $Q_P^{-1} \pm \Delta Q_P^{-1}$ | $Q_S^{-1} \pm \Delta Q_S^{-1}$ | $Q_P^{-1}/Q_S^{-1}$ |
|---|---|---|---|
| 1.5 | (13.8±3.3)× $10^{-3}$ | (6.0±1.4) × $10^{-3}$ | 2.30 |
| 3.5 | (4.9±1.5) × $10^{-3}$ | (1.8±0.3) × $10^{-3}$ | 2.72 |
| 6 | (2.3±0.5) × $10^{-3}$ | (0.9±0.2) × $10^{-3}$ | 2.55 |
| 9 | (1.3±0.3) × $10^{-3}$ | (0.56±0.15) × $10^{-3}$ | 2.32 |
| 12 | (0.9±0.2) × $10^{-3}$ | (0.4±0.09) × $10^{-3}$ | 2.25 |

**Table** 6:

| | $Q_P^{-1}$ | $Q_S^{-1}$ |
|---|---|---|
| Western Nagano, Japan (Yoshimoto et. al. 1998) | $52 \times 10^{-3} f^{(-0.66)}$ | $34 \times 10^{-3} f^{(-0.12)}$ |
| Baoshan, China (Qin-cai et al. 2005) | $11.55 \times 10^{-3} f^{(-0.93)}$ | $8.67 \times 10^{-3} f^{(-0.86)}$ |
| East-Central Iran (Ma'hood et al. 2009) | $25 \times 10^{-3} f^{(-0.99)}$ | $19 \times 10^{-3} f^{(-1.02)}$ |
| Kanto, Central Japan (Yoshimoto et. al. 1993) | $31 \times 10^{-3} f^{(-0.95)}$ | $12 \times 10^{-3} f^{(-0.73)}$ |
| Zarand (Ma'hood et al. 2009) | $28 \times 10^{-3} f^{(-1.03)}$ | $17 \times 10^{-3} f^{(-1.0)}$ |
| South Korea (Chung and Sato, 2001) | $9 \times 10^{-3} f^{(-1.05)}$ | $4 \times 10^{-3} f^{(-0.70)}$ |
| Kachchh, Gujarat (Sharma et. al., 2008) | $13 \times 10^{-3} f^{(-0.87)}$ | $10 \times 10^{-3} f^{(-0.86)}$ |
| Kopili Region (This Study) | $(23.8 \pm 6) \times 10^{-3} f^{(-1.2 \pm 0.008)}$ | $(10.2 \pm 2) \times 10^{-3} f^{(-1.3 \pm 0.02)}$ |